\documentclass[sort&compress, review]{elsarticle}
\journal{Physics Letters B}
\usepackage[utf8]{inputenc}
\usepackage[english]{babel}
\usepackage[fleqn]{amsmath}
\usepackage{amsfonts}
\usepackage{amssymb}
\usepackage{url,xcolor,slashed,bbm}
\usepackage{siunitx}
\usepackage{xifthen}
\usepackage{multirow,listliketab,tabu,longtable}
\usepackage{array}
\usepackage[expansion=true] {microtype}	
\usepackage{mathtools,colonequals,bm,rotating}
\usepackage{etextools}
\usepackage{graphicx,picture,placeins}
\usepackage[caption=false]{subfig}
\usepackage{multirow}
\usepackage[hypertexnames=false]{hyperref}
\hypersetup{
  colorlinks = true,
  urlcolor   = black,
  linkcolor  = black,
  citecolor  = black
}
\usepackage[color]{changebar}
\cbcolor{red}
\graphicspath{{./Bilder/}}
\newcommand*{\eh}[1]{\mathrm e^{#1}}

\newcommand*{\diff}{\mathrm d}

\newcommand*{\lint}{\int\limits}

\newcommand{\eperp}{{\epsilon_\perp}}
\newcommand{\Ell}{\mathrm{E}}
\renewcommand{\vec}[1]{{\bm #1}}
\renewcommand{\Re}{\operatorname{Re}}

\allowdisplaybreaks

\begin{document}

\begin{frontmatter}

\title{Lifting shell structures in the dynamically assisted Schwinger effect in periodic fields}

\author[hzdr,itp]{A.~Otto}
\ead{a.otto@hzdr.de}
\author[jena]{D.~Seipt}
\author[wroclaw]{D.~Blaschke}
\author[hzdr,itp]{B.~K\"ampfer}
\author[saratov]{S.A.~Smolyansky}

\address[hzdr]{Institute of Radiation Physics, Helmholtz-Zentrum Dresden-Rossendorf,\\Bautzner Landstra\ss e 400, 01328 Dresden, Germany}
\address[itp]{Institut f\"ur Theoretische Physik, Technische Universit\"at Dresden,\\Zellescher Weg 17, 01062 Dresden, Germany}
\address[jena]{Helmholtz-Institut Jena, Fr\"obelstieg 3, 07743 Jena, Germany}
\address[wroclaw]{Group of Elementary Particle Theory, Institute for Theoretical Physics,\\University of Wroclaw, pl. M. Borna 9, 50-204 Wroclaw, Poland}
\address[saratov]{Department of Physics, Saratov State University, 410071 Saratov, Russia}

\begin{abstract}
The dynamically assisted pair creation (Schwinger effect) is considered for the superposition of two periodic electric fields acting a finite time interval.
We find a strong enhancement by orders of magnitude caused by a weak field with a frequency being a multitude of the strong-field frequency.
The strong low-frequency field leads to shell structures which are lifted by the weaker high-frequency field.
The resonance type amplification refers to a new, monotonously increasing mode, often hidden in some strong oscillatory transient background which disappears during the smoothly switching off the background fields, thus leaving a pronounced residual shell structure in phase space.
\end{abstract}

\end{frontmatter}

\section{Introduction}
For many decades the Schwinger effect \cite{schwinger} has been considered crucial for testing non-perturbative QED as a pillar of the standard model of particle physics in the strong-field regime.
An obvious motivation for the broad interest can be seen in the formal structure and numerical smallness of the decay rate $R$ of a static, purely electric field $E_0$ into a state with on-shell electrons and positrons which screen the original field.
Schwinger's seminal formula was $R\propto E_0^2\exp\left(-\pi E_c/E_0\right)$ in leading order, where the scale is set by the electron's mass $m$ and charge $e$ reading $E_c=m^2/e$ (we use units with $\hbar=c=1$) first introduced by Sauter \cite{sauter}.
Presently achievable long-living fields in the laboratory are weak compared to $E_c$, $E_0\ll E_c$.
Accordingly, the Schwinger rate is exponentially small and has escaped an experimental verification until now.

The fields created in peripheral relativistic heavy-ion collisions are short-lived, of the order of a few $\SI{}{fm}/c$~\cite{basar_magneto-sono-luminescence_2014}, thus not suitable for an exploration of the original Schwinger effect which is for a spatio-temporal constant field.
Nevertheless, a plethora of interesting strong-field effects are under consideration \cite{tuchin_particle_2013}.
For instance, magnetars are astrophysical objects with strong fields which could serve for identifying Schwinger type effects \cite{kouveliotou_x-ray_1998, mereghetti_strongest_2008}.
One should also recall that the Schwinger effect for chromoelectric fields is employed in phenomenological models of particle production in strong interaction processes \cite{casher_chromoelectric-flux-tube_1979, roberts_dyson-schwinger_2000, prozorkevich_vacuum_2004}.

Two further aspects highlight the role of the Schwinger effect.
(i) It is conceivable that QED is an effective weak-field theory which breaks down for fields of the order of $E_c$. 
(ii) A long-living field $\mathcal O(E_c)$ can not be achieved due to screening processes and cascades which consume and transfer the original field energy into other degrees of freedom, as discussed in \cite{fedotov_limitations_2010, bell_possibility_2008, elkina_qed_2011, nerush_laser_2011}. We mention further that the decay of a strong external field due to particle production is not a privilege of QED, but is generic. For instance, the Hawking radiation off a horizon is a famous example w.r.t.\ gravitational fields \cite{birrell_davies, mamayev_mostepanenko}.

In the course of seeking set-ups which could offer the opportunity to verify the above static Schwinger effect, the idea has been explored that ultra-intense laser fields could enable the detection of the dynamical Schwinger effect \cite{brezin_pair_1970}.
For instance, in the antinodes of two counter propagating, linearly polarized laser beams we have a periodic (frequency $\nu$), essentially electric field $E(t)$ with spatial homogeneity length of $\mathcal O(1/\nu)$ which is, for optical lasers, much larger than the Compton wave length $\lambda_C=2\pi/m$ of the electron.
The prospects of $e^+e^-$ pair production in dependence on $E_0$ and $\nu$ have recently been analyzed
\cite{blaschke_properties_2013}.
While in a plane wave or null field the pair production rate is zero \cite{schwinger}, a focused laser field provides a non-zero rate, as pointed out in \cite{narozhny_pair_2004}.
However, the rate is still very small, unless such ultra-intense laser fields as envisaged at ELI \cite{ELI} are at our disposal.
Finally we mention Ref.~\cite{dreisow_vacuum_2012}, where the mimicking of the dynamical Schwinger effect is accomplished in an all-optics setup of a wave guide with curved optical axis.

While the Schwinger effect is originally related to a tunneling process, which escapes the standard perturbative QED described by Feynman diagrams, in the dynamically assisted Schwinger effect \cite{schutzhold_dynamically_2008, fey_momentum_2012, dunne_catalysis_2009} the tunneling is combined with a multi-photon process, thus potentially enhancing the pair production rate significantly.
The essence is a combination of a strong field (may be slowly varying) with a weak field which introduces, in particular, a high-frequency component.
Various combinations have recently been investigated to look for optimum parameter settings.
In Refs. \cite{orthaber_momentum_2011, kohlfurst_optimizing_2013}, the superposition of two Sauter pulses was considered;
Ref. \cite{sicking_bachelor_2012} analyzed the superposition of a strong Sauter pulse with various other weak-pulse shapes.
The Sauter pulse has a d.c.\ component and can hardly be shaped with present laser technologies.
It is therefore tempting to investigate the rate enhancement in the superposition of two periodic fields, e.g.\ as recently done also in~\cite{hebenstreit_optimization_2014, akal_electron-positron_2014}.
Such a situation seems to be more realistic in respect to a suitable combination of XFEL and optical laser beams.
The opportunities at plain XFEL beams are considered in \cite{ringwald_pair_2001}.
References~\cite{nuriman_enhanced_2012, oluk_electron-positron_2014} consider the frozen-out early-time population of low-momentum electrons (positrons) in various field configurations, while we consider the residual phase space occupation with a realistic (smooth) switching on/off the combined fields.

Our framework is the kinetic equation for the single-particle distribution derived in \cite{schmidt_quantum_1998}, see also \cite{blaschke_pair_2006, blaschke_properties_2013, blaschke_influence_2013, smolyansky_photon_2012}.
Despite the ostensible simplicity of the kinetic equation and the possibility to give a compact expression for its solution, it is fairly intransparent due to the non-linear and non-Markovian character.
Therefore, it is hardly possible to read off in a simple manner the dependence of the solution on the field parameters.
WKB type approaches \cite{kleinert_electron-positron_2008, dunne_catalysis_2009, di_piazza_pair_2004}, the world line formalism \cite{gies_pair_2005} and optimization theory \cite{kohlfurst_optimizing_2013} have been developed to gain further insights into the pair production process.
We here rely on numerical solutions of the kinetic equation to elucidate parameter regions where the dynamically assisted Schwinger effect in two periodic fields, which are smoothly switched on and off, leads to a significant enhancement of the rate.
The numerical  simulations (section~\ref{sec:ke}) are accompanied and interpreted by analytical approximations (section~\ref{sec:shell_structure_shape}) explaining the shell structure in phase space.
This is supplemented by a systematic scan of parameter dependence (section~\ref{sec:param}).
Our summary is given in section~\ref{sec:summary}.

\section{Solutions of quantum kinetic equations\label{sec:ke}}
The quantum-kinetic equation without back reaction for the time ($t$) evolution of the one-particle distribution function $f$ (cf.~\cite{dabrowski_super-adiabatic_2014} for a discussion of the meaning of $f$) summed over spin projections is given either as an integro-differential equation \cite{schmidt_quantum_1998, blaschke_properties_2013, hebenstreit_diss_2011}
\begin{align}
\dot f(\vec p, t) = Q(\vec p, t)\lint_{t_0}^t\!\!\diff t'\,Q(\vec p, t')\bigl[1 - \eta f(\vec p, t')\bigr]\cos 2\bigl[\Theta(\vec p, t) - \Theta(\vec p, t')\bigr]
\label{eq:integro}
\end{align} 
or equivalently as a system of three coupled differential equations
\begin{subequations}
\begin{align}
\dot f(\vec p, t) &= Q(\vec p, t)\,u(\vec p, t)\:, \label{eq:ke1}\\
\dot u(\vec p, t) &= Q(\vec p, t)\,[1-\eta f(\vec p, t)] - 2\omega(\vec p, t)\,v(\vec p, t)\:,  \label{eq:ke2}\\
\dot v(\vec p, t) &= 2\omega(\vec p, t)\,u(\vec p, t)\:,
\end{align}
\label{eq:ke}
\end{subequations}
where $u$ and $v$ denote auxiliary quantities and $\Theta$, $\omega$ and $Q$ are defined by
\begin{align}
\Theta(\vec p, t) &= \lint_{t_0}^t\!\!\diff t'\,\omega(\vec p, t')\:,\\
\label{eq:Theta}
\omega(\vec p, t) &= \sqrt{\epsilon_\perp^2+\bigl(p_\parallel-eA(t)\bigr)^2}\:,\\
Q(\vec p, t) &= \frac{eE(t)\eperp}{\omega^2(\vec p, t)}\:,
\end{align}
with $A(t)$ and $E(t)=-\dot A(t)$ being the $z$ component of the vector potential and the electric field, respectively.
Our field is thus assumed spatially homogeneous, pointing along the $z$ direction.
Consequently, $p_\parallel$ denotes the momentum (e.g. of electrons) parallel to the $z$ axis and $p_\perp$ the momentum perpendicular to it;
$\eperp=\sqrt{m^2+p_\perp^2}$ is the transverse energy;
$p_\parallel$ and $p_\perp$ are components of the three-vector $\vec p$.
From here on, we set $t_0 = 0$ and employ the initial conditions $f(t_0) = u(t_0) = v(t_0) = 0$.
The parameter $\eta$ in~\eqref{eq:integro} and~\eqref{eq:ke2} distinguishes the full solution ($\eta=1$, considered in this section) from the low-density approximation ($\eta=0$, employed in section~\ref{sec:shell_structure_shape}).

In what follows we consider the synchronized superposition of a slow strong field (``$1$'') and a fast weak field (``$2$'') with potential
\begin{align}
A(t) = \left(\frac{E_1}{\nu}\cos(\nu t) + \frac{E_2}{N\nu}\cos(N\nu t)\right) K(\nu t)
\label{eq:A_field}
\end{align}
where $\nu=2\pi/T$ is the frequency of the slow field and $N$ the ratio of the frequencies chosen to be integer.
We utilize a $C^\infty$ envelope function (which is infinitely often differentiable)
\begin{align}
K(\tau) = \begin{cases}
\text{$0$ for $\tau<0$,} &\\
\text{smooth transition for $0<\tau<\tau_\text{ramp}$,} &\\
\text{$1$ for $\tau_\text{ramp}<\tau<\tau_\text{ramp}+\tau_\text{f.t.}$,} &\\
\text{smooth transition for $\tau_\text{ramp}+\tau_\text{f.t.}<\tau<\tau_\text{pulse}$,} &\\
\text{$0$ for $\tau_\text{pulse}<\tau$,} &
\end{cases}
\end{align}
which is chosen as $K(\tau) = h\left( \frac{\tau}{\tau_\text{ramp}} \right)\, h\left( \frac{\tau_\text{pulse}-\tau}{\tau_\text{ramp}} \right)$ where $h(x) = \frac{g(x)}{g(x)+g(1-x)}$ and $g(x) = 0$ for $x \le 0$, $g(x) = \eh{-\frac{1}{x}}$ for $x > 0$.
The field~\eqref{eq:A_field} is therefore smoothly switched on and off for a suitable choice of the ramping (``ramp'') interval from $0$ to $\tau_\text{ramp}$ and deramping interval from $\tau_\text{f.t.}+\tau_\text{ramp}$ to $\tau_\text{pulse} = \tau_\text{f.t.}+2\tau_\text{ramp}$;
the flat-top (``f.t.'') interval is from $\tau_\text{ramp}$ to $\tau_\text{f.t.}+\tau_\text{ramp}$.
The potential~\eqref{eq:A_field} and thus also the electric field acts for the finite duration $\tau_\text{pulse}$.
We have chosen $\tau_\text{ramp} = 5\cdot 2\pi$ and $\tau_\text{f.t.} = 50\cdot 2\pi$ meaning five (fifty) oscillations of field ``1'' for ramping and deramping (the flat-top interval).
Thus, the field configuration~\eqref{eq:A_field} is a special model for the spatial homogeneity region of a common antinode of several (at least four) pair-wise counterpropagating synchronized beams.
In the present study we focus on time scales and field strengths similar to those in \cite{orthaber_momentum_2011}: $E_1 = \num{0.1}E_c$ and $\nu = \num{0.02}m$, $E_2 = 0\ldots\num{0.05}E_c$ and $N = 10\ldots 50$.
That means the individual Keldysh parameters are $\gamma_1 = (E_c/E_1)(\nu/m) = \num{0.2}$ and $\gamma_2 = (E_c/E_2)(N\nu/m) = \mathcal O(4\dots\infty)$.
While this parameter regime does not exactly match presently available XFEL and intense laser technology, it allows for an easy numerical treatment of the kinetic equations (and comparison with available literature).
In~\cite{brezin_pair_1970}, $\gamma\ll1$ is referred to as tunneling regime, while $\gamma\gg1$ is the multi-photon regime.

\begin{sidewaysfigure}
\centering
\includegraphics[width=\textwidth]{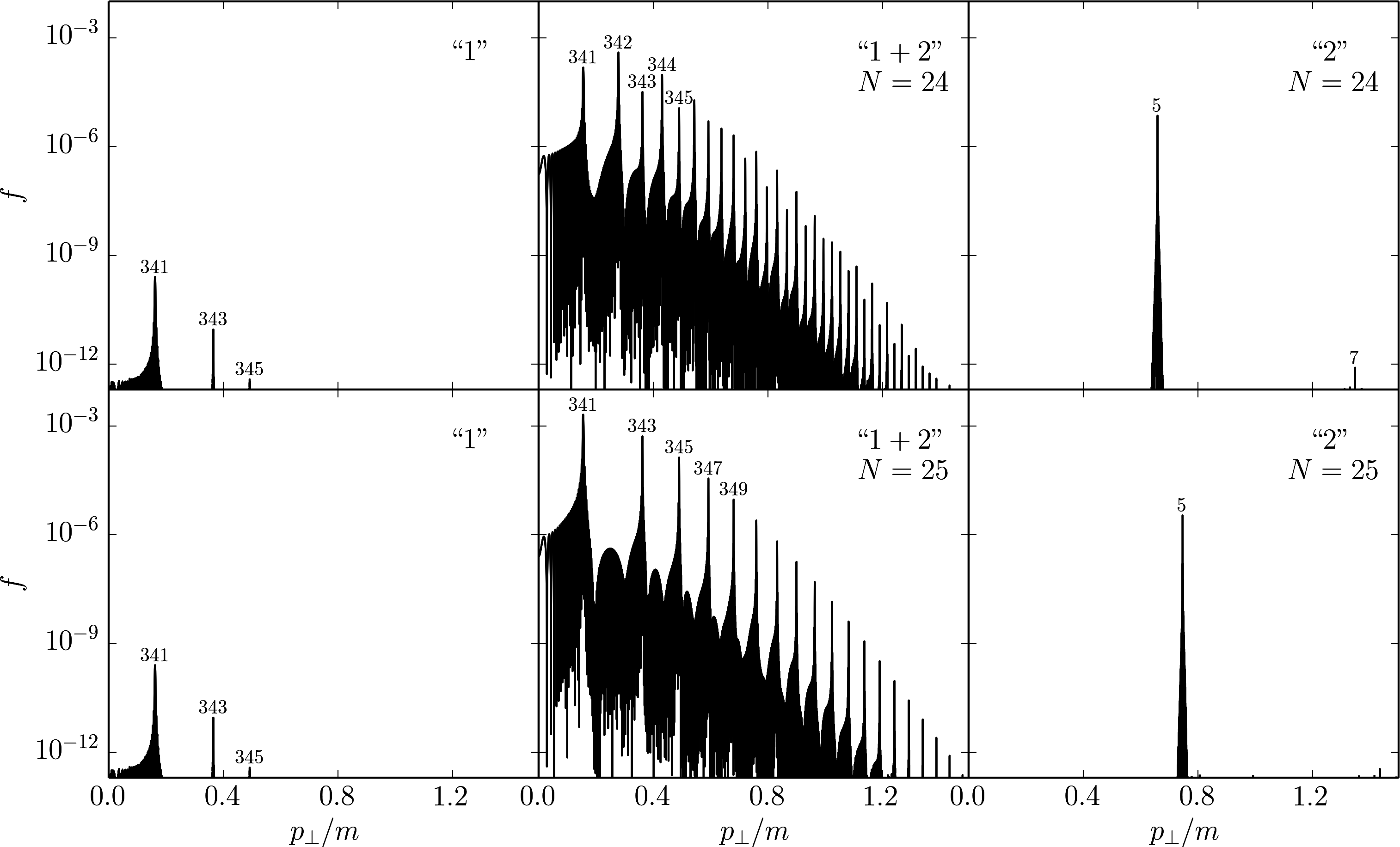}
\caption{The residual phase space distribution $f(p_\perp, p_\parallel=0)$ as a function of $p_\perp$.
The parameters are $E_1=\num{0.1}E_c$, $\nu=\num{0.02}m$, $E_2=\num{0.05}E_c$, $\tau_\text{ramp}=5\cdot 2\pi$, $\tau_\text{f.t.}=50\cdot 2\pi$.
Left panels: the slow strong field alone (top and bottom panels are the same, but are displayed for an easy comparison with middle and right panels); middle panels: for the field~\eqref{eq:A_field}; right panels: the fast weak field alone.
Top: $N=24$; bottom: $N=25$.
The labels are shell numbers $\ell$ according to~\eqref{eq:shell_condition}.
Shell number $7$ in the bottom right panel is outside the displayed region due to the increase of $N$.}
\label{fig:cut}
\end{sidewaysfigure}

Solutions of~\eqref{eq:ke1} for $\eta = 1$ (i.e.\ with Pauli blocking) and for $p_\parallel = 0$ are exhibited in Fig.~\ref{fig:cut} for $\nu t > \tau_\text{pulse}$ where, according to~\eqref{eq:ke}, $\dot f = 0$ since $E(\nu t > \tau_\text{pulse}) = 0$.
(That means, $f(\nu t>\tau_\text{pulse})$ represents the residual phase space distribution within the considered framework.)
The middle panels in Fig.~\ref{fig:cut} exhibit the residual phase space distributions in $p_\perp$ direction at $p_\parallel=0$ for the field~\eqref{eq:A_field}, while the left (right) panels are for the strong (weak) field alone.
One observes pronounced peaks which continue (albeit at different positions) when displaying other cuts in the $p_\perp$-$p_\parallel$ plane or sharp ridges in contours over the $p_\perp$-$p_\parallel$ plane.
These peaks or ridges are referred to as shell structures, already described, for a single periodic field, in various previous papers~\cite{blaschke_properties_2013, alkofer_pair_2001, mocken_nonperturbative_2010, ruf_pair_2009}, originally found in~\cite{brezin_pair_1970} and further elaborated in~\cite{popov_resonant_1973, narozhny_pair_1973, mostepanenko_1974, popov_1974}.
From Fig.~\ref{fig:cut} one infers that the residual phase space occupations for any one of the two field contributions that appear in~\eqref{eq:A_field} are much smaller than the phase space occupations for the superposition of both fields.
For instance, shell $\ell_{(\text{``1''})}= 341$ (left panels in Fig.~\ref{fig:cut}) with peak altitude \num[round-mode=places,round-precision=1]{2.53171817e-10} becomes, due to the impact of the field ``2'', shell $\ell_{(\text{``1''+``2''})} = 341$ (middle panels in Fig.~\ref{fig:cut}) with peak altitude \num[round-mode=places,round-precision=1]{1.50507358e-04} or \num[round-mode=places,round-precision=1]{2.02808343e-03} depending on $N$.
The peak pattern is dominated by the slow strong field ``1'', where ``2'' lets even shells additionally appear, e.g.\ shells $\ell_{(\text{``1''+``2''})} = 342$, $344$ etc.\ for $N=25$, which are not visible for the field ``1'' alone (cf.\ left panels).
Due to the comparatively high frequency $\nu_2 = N\nu_1$ of the field ``2'', the shell numbers $\ell_{(\text{``2''})}$ are much smaller and the corresponding peaks are much higher, but individual structures resembling the right panels in Fig.~\ref{fig:cut} are not evident in the middle panels.
The assistance of field ``2'' consists obviously in lifting the pattern governed by field ``1''.

The found non-linear amplification is huge -- much larger than for the superposition of two Sauter pulses in~\cite{orthaber_momentum_2011}.
References~\cite{nuriman_enhanced_2012, oluk_electron-positron_2014} also report very strong amplification effects for periodic fields, but for a very special shape function $K$ and a different early-time mode.
Other field configurations are considered in~\cite{hebenstreit_momentum_2009, jiang_enhanced_2013, ren_pair_2012}, where relatively strong effects in the momentum dependence and particle rate are found by modifying a Gaussian electric field by a subcycle sinusoidal field.

\section{Shell structure and shell shape\label{sec:shell_structure_shape}}
To arrive at a qualitative understanding of the numerical results of the previous section we resort to the low-density approximation (exponentiating results in the Markovian approximation~\cite{schmidt_non-markovian_1999})
\begin{align}
f(\vec p, t) &= \frac{1}{2}\left| I(\vec p, t) \right|^2 \: , \label{eq:f_from_I}\\
I(\vec p, t) &= \lint_0^t\!\!\diff t' \, \frac{eE(t')\eperp}{\omega(\vec p, t')^2} \, \eh{2i\Theta(\vec p, t')} \label{eq:low-density}
\end{align}
which discards the Pauli blocking by setting $\eta = 0$ in~\eqref{eq:integro} and~\eqref{eq:ke} or $f \ll 1$ in~\eqref{eq:integro}.
While asymptotically $f \ll 1$ in Fig.~\ref{fig:cut}, at intermediate times this needs not necessarily be the case.
Nevertheless, the low-density approximation yields sufficiently accurate results (on the percent level) within the considered parameter domain w.r.t.\ shell positions, peak heights and widths provided by the following harmonic analysis.

\subsection{Shell structure}
Given the periodicity of $\omega(\vec p, t)$ w.r.t.\ to $T$ when considering $K=1$, a Fourier representation of~\eqref{eq:Theta} is in order~\cite{brezin_pair_1970, blaschke_influence_2013}:
\begin{align}
\Theta(\vec p, t) = \Omega(\vec p) t + P(\vec p, t)\:,
\end{align}
where $\Omega(\vec p) = \frac{1}{T}\int_0^T\!\diff t\,\omega(\vec p, t)$ is the Fourier zero-mode (called `renormalized frequency' in~\cite{brezin_pair_1970}) and $P(\vec p, t)$ is a $T$-periodic function.
The resulting expression $I(\vec p, t) = \int_0^{t}\!\diff t' F(\vec p, t')\eh{2i\Omega(\vec p) t'}$ with the $T$-periodic function
\begin{align}
F(\vec p, t) = \frac{\eperp eE(t)}{\omega(\vec p, t)^2}\eh{2iP(\vec p, t)}
\end{align}
calls for a second Fourier expansion $F(\vec p, t) = \sum_\ell F_\ell(\vec p)\,\eh{-i\ell\nu t}$ with the Fourier coefficients
\begin{align}
F_\ell(\vec p) = \frac{1}{T}\lint_0^T\!\! \diff t\, F(\vec p, t)\eh{i\ell\nu t}\:.
\label{eq:fourier_coefficients}
\end{align}
Due to the symmetry of the functions $\omega(t)$ and $\cos 2\Theta(t)$ and antisymmetry of $E(t)$ and $\sin 2\Theta(t)$ w.r.t.\ $t=T/2$, one finds $\Re F_\ell = 0$, which can be used to check the accuracy of numerical calculations.
Upon time integration in~\eqref{eq:low-density} one gets
\begin{align}
I(\vec p, t) = \sum_\ell iF_\ell(\vec p)\frac{\eh{-i(\ell\nu-2\Omega(\vec p))t}-1}{\ell\nu-2\Omega(\vec p)} \:,
\label{eq:I_fourier_integrated}
\end{align}
indicating that for
\begin{align}
\ell\nu - 2\Omega(\vec p) = 0
\label{eq:shell_condition}
\end{align}
sharp ridges/peaks can appear in the distribution function.
Solutions of~\eqref{eq:shell_condition} are, for a given value of $\ell$ which we call shell number, $p_\perp^{(\ell)}(p_\parallel)$ or, for $p_\parallel = 0$, simply $p_\perp^{(\ell)}$.
(The labels in Fig.~\ref{fig:cut} are just these shell numbers $\ell$.)
The small-momentum expansion of $\Omega(\vec p)$ reads 
\begin{align}
\Omega(\vec p) = \Omega(\vec p=0) + \Omega_\parallel p_\parallel + \Omega_1p_\perp^2 + \Omega_2p_\parallel^2
\end{align}
with
$\Omega_\parallel = -T^{-1}\int_0^T\diff t\,eA(t)/\omega(t,\vec p=0)$,
$\Omega_1 = (1 + \gamma_1\partial/\partial\gamma_1 + \gamma_2\partial/\partial\gamma_2)\Omega(\vec p=0)$,
$\Omega_2 = (1 - \gamma_1\partial/\partial\gamma_1 - \gamma_2\partial/\partial\gamma_2)\Omega_1$.
The limit $\gamma_2\rightarrow\infty$ or $E_2\rightarrow0$ recovers~\cite{popov_resonant_1973} with $\Omega_\parallel=0$.

The leading-order behaviour of $\Omega(\vec p=0)$, which also depends on the parameters $\gamma_1$, $\gamma_2$ and $N$, is for $\gamma_1/\gamma_2\ll1$ given by $(2m/\pi) \sqrt{1+1/\gamma_1^2}\,\Ell \bigl(1/(1+\gamma_1^2)\bigr)$~\cite{popov_resonant_1973}, where $\Ell(x)$ is the complete elliptic integral with $\Ell(0) = \pi/2$ and $\Ell(1) = 1$, i.e.\ $\left.\Omega(\vec p=0)\right|_{\gamma_1\rightarrow0}\rightarrow 2m/(\gamma_1\pi)$ and $\left.\Omega(\vec p=0)\right|_{\gamma_1\rightarrow\infty}\rightarrow m$ implying $\Omega(\vec p=0) > m$.
(The corrections to the leading-order term are small, e.g.\ $<\num{0.1}\%$ for $\gamma_1=\num{0.2}$, $\gamma_2\ge10$ and $N\ge10$, with signs depending on $\gamma_2$ and $N$.)
Numerically, the effective mass $m_* = m\sqrt{1+1/(2\gamma_1^2)}$~\cite{schmidt_non-markovian_1999} agrees with $\Omega(\vec p =0)$ better than $1\%$ ($\num{7.3}\%$) for $\gamma_1\ge 1$ ($\ge\num{0.2}$) and $\gamma_1/\gamma_2\ll 1$.
Towards the tunneling regime, i.e.\ at smaller values of $\gamma_1$, the effective mass concept is found in~\cite{kohlfurst_effective_2014} to be less adequate and one could argue that $\Omega(\vec p)$ is a more sensible quantity, e.g. for identifying shell positions $\vec p^{(\ell)}$.
Since $\Omega(\vec p)$ increases with increasing field strength $E_1$ at fixed frequency and large values of $\gamma_2$, the previously lowest shell, characterized by $\ell_\text{min}\nu$, can ``disappear'' if $\Omega(\vec p=0)$ becomes larger than $\ell_\text{min}\nu$.
This is the analog of channel closing in atomic ionization (ATI).

\subsection{On-shell occupancy\label{sec:on-shell_occupancy}}
On shell $\ell$, \eqref{eq:I_fourier_integrated} inserted in~\eqref{eq:f_from_I} delivers
\begin{align}
f(\vec p^{(\ell)}, t) &= \frac{1}{2}\left| iF_l(\vec p^{(\ell)})t + \sum_{k \ne \ell} iF_k(\vec p^{(\ell)}) \frac{\eh{i(k\nu-2\Omega(\vec p^{(\ell)}))t}-1}{k\nu-2\Omega(\vec p^{(\ell)})} \right|^2\nonumber\\
&= \frac{1}{2}\left|F_\ell(\vec p^{(\ell)})\right|^2t^2 + G(\vec p^{(\ell)}, t)t + H(\vec p^{(\ell)}, t),
\label{eq:f_on_shell}
\end{align}
where $G(\vec p^{(\ell)}, t)$ and $H(\vec p^{(\ell)}, t)$ are bounded oscillating functions depending on $\vec p^{(\ell)}$.
The peak height of a shell at position $\vec p^{(\ell)}$ increases accordingly quadratically with time (first term in~\eqref{eq:f_on_shell}), being periodically modulated with a linearly increasing (second term) and a constant amplitude (last term).
Due to the superposition of these modes the actual transient time evolution can be quite involved but lacks a physical meaning, as recalled in~\cite{dabrowski_super-adiabatic_2014}.
We observed in our numerical simulations based on~\eqref{eq:ke}, however, that after smoothly switching off the field, the peak height $f(\vec p^{(\ell)}, \nu t>\tau_\text{pulse})$ coincides with the first term in~\eqref{eq:f_on_shell}:
The numerical evaluation of $F_\ell(\vec p^{(\ell)})$ according to~\eqref{eq:fourier_coefficients} and using it in $f(\vec p^{(\ell)}, \nu t>\tau_\text{pulse}) = \frac{1}{2} | F_\ell(\vec p^{(\ell)}) |^2t_\text{f.t.}^2$ with $t_\text{f.t.}$ as flat-top interval time agrees well with numerical results of the peak heights by integrating~\eqref{eq:ke}.
Thus $\frac{1}{2} | F_\ell(\vec p^{(\ell)}) |^2t_\text{f.t.}^2$ can be identified with the residual on-shell occupancy $f(p^{(\ell)})$.

\subsection{Shell shape}
For a more detailed account of the shell shape, let us expand~\eqref{eq:I_fourier_integrated} for $p_\parallel = 0$ around $p_\perp^{(\ell)}$ by setting $p_\perp = p_\perp^{(\ell)} + \Delta p$ to find in leading order of $\Delta p$
\begin{align}
f(p_\perp^{(\ell)} + \Delta p, 0, t) \approx \frac{1}{2} \left| F_\ell(p_\perp^{(\ell)}, 0) \right|^2 \frac{\sin^2\left( \Omega'(p_\perp^{(\ell)}, 0)\Delta p\, t \right)}{\left( \Omega'(p_\perp^{(\ell)}, 0)\Delta p \right)^2}\:.
\label{eq:shell_profile}
\end{align}
Since the full width at half maximum (FWHM) of $\sin^2(xt)/x^2$ evolves as $\propto 1/t$, the FWHM of $f(p_\perp^{(\ell)} + \Delta p, 0, t)$ evolves as $\propto \left( \Omega'(p_\perp^{(\ell)}, 0)\,t \right)^{-1}$, i.e.\ the important result arises that the shell width shrinks with time.
(Here, $\Omega' = \partial\Omega/\partial p_\perp$ is the slope of $\Omega(p_\perp, 0)$ at shell position $p_\perp^{(\ell)}$.)
The transverse momentum integral for the contribution of the shell $\ell$ can be estimated by
\begin{align}
\lint_0^\infty\!\! \diff p_\perp p_\perp f(p_\perp^{(\ell)}+\Delta p, 0, t) \approx \frac{\pi}{2}\frac{p_\perp^{(\ell)}}{\left| \Omega'(p_\perp^{(\ell)}, 0) \right|}\left| F_\ell(p_\perp^{(\ell)}, 0) \right|^2 t\:,
\label{eq:shell_line_contribution}
\end{align}
i.e.\ despite the quadratic growth of the shell height, the shrinking causes a linear increase with time of the line integrated density.
In fact, the residual density is determined by~\eqref{eq:shell_line_contribution} with $t\rightarrow t_\text{f.t.}$, as our numerical investigations based on~\eqref{eq:ke} show.
Neglecting the pedestrials under the sharp peaks (cf.\ Fig.~\ref{fig:cut}) the residual density $n=2\pi\int\!\!\diff p_\parallel\diff p_\perp p_\perp f(\vec p)$ can be estimated by summing over all shells $\ell\ge\ell_\text{min}$, i.e.
\begin{align}
n \approx 2\pi^2 \sum_{\ell=\ell_\text{min}}^\infty \frac{{p_\perp^{(\ell)}}^2}{|\Omega'(p_\perp^{(\ell)}, 0)|}\,| F_\ell(p_\perp^{(\ell)}, 0)|^2\, t_\text{f.t.}
\label{eq:density_estimator}
\end{align}
when neglecting the anisotropy in phase space by setting ${p_\perp^{(\ell)}}^2 + {p_\parallel^{(\ell)}}^2 = {p_\perp^{(\ell)}}^2(p_\parallel=0)$ and the peculiarities for $p_\parallel\ne0$.
Numbers are discussed in the Appendix.

\section{Survey on the parameter dependence \label{sec:param}}
\begin{figure}
\centering
\includegraphics[width=\textwidth]{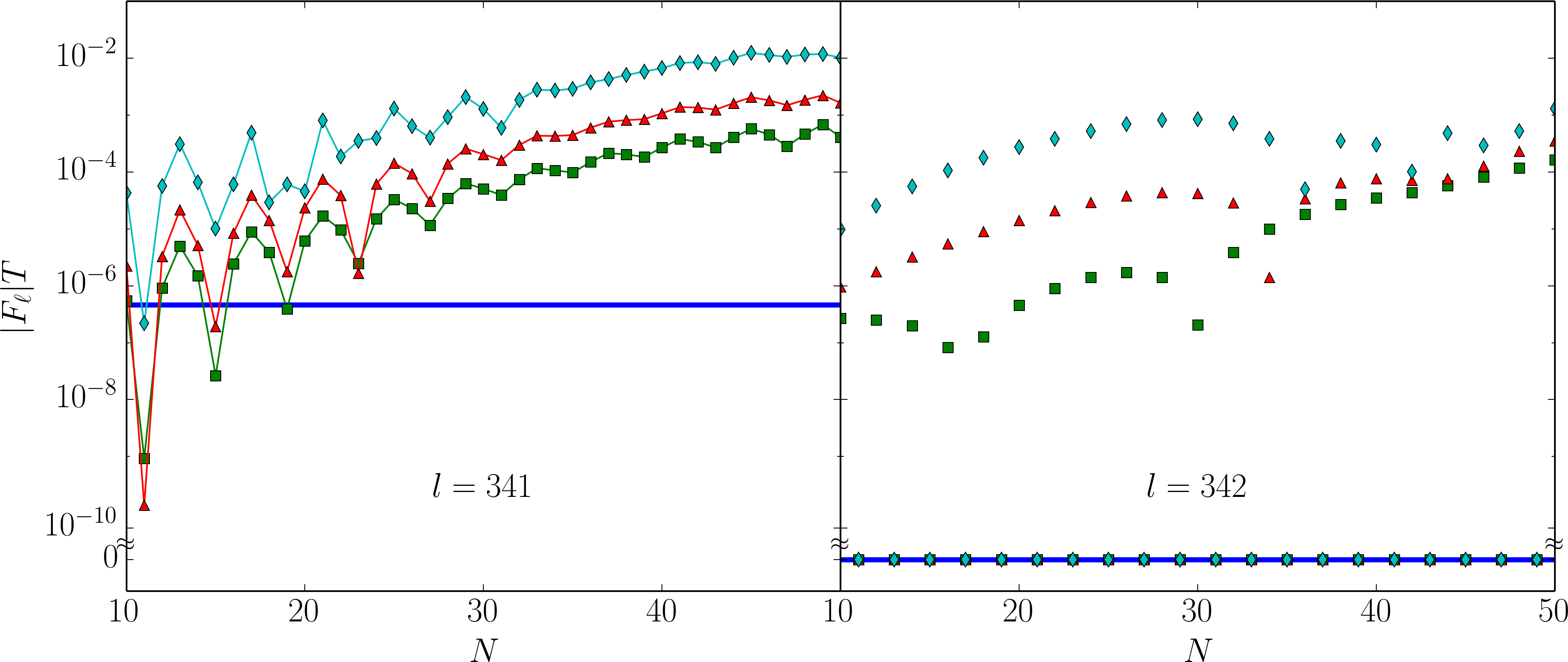}
\caption{Fourier coefficients $|F_\ell|$ as a function of $N$ for $p_\parallel=0$ and shells $\ell=341$ (left, lines are drawn to guide the eyes; the symbols depict the results for integer values of $N$) and $\ell=342$ (right) and various field intensities $E_2$ (green squares: $E_2=\num{0.01}E_c$, red triangles: $E_2=\num{0.02}E_c$, cyan diamonds: $E_2=\num{0.05}E_c$).
The blue lines are for $E_2=0$, i.e.\ the field ``1'' alone.
Note the $\Delta N=2$ staggering for the even shell (right).}
\label{fig:fourier_coefficients_N}
\end{figure}

After having identified the decisive role of the Fourier coefficients $F_\ell$ defined in~\eqref{eq:fourier_coefficients} for shell heights and widths and residual density we proceed with a brief survey on some systematics.
Figure~\ref{fig:fourier_coefficients_N} exhibits the Fourier coefficients for shells $\ell=341$ and $\ell=342$ which are the lowest allowed shells for both the field~\eqref{eq:A_field} (cf. middle column in Fig.~\ref{fig:cut}) and the slow strong field alone (cf. left column in Fig.~\ref{fig:cut}).
Let us first consider shell $341$.
One observes for sufficiently large values of $N$ and field strength $E_2$ of the fast weak field a strong increase due to the action of the faster field.
The blue line is for the slow strong field alone, i.e.\ $E_2=0$, meaning that all points above indicate an amplification by the fast weak field.
(Remember that the density accumulated in the shells, according to~\eqref{eq:shell_profile}, is proportional to $|F_\ell|^2$.)
For $N > 30$ the apparent $\Delta N = 4$ periodicity dies out, and $|F_{341}|$ grows with increasing $N$ and $E_2$.
Since the dynamical phase $\Theta(t)$ introduces a highly oscillating part of the integrand in~\eqref{eq:fourier_coefficients}, small ``detunings'' by variations of $N$ and $E_2$ might cause the irregularly looking pattern at smaller values of $N$, where the impact of the second field can induce even a depletion of shell occupancy.
The pattern exhibited in the left panel of Fig.~\ref{fig:fourier_coefficients_N} continues to higher shells with odd $\ell$, however with decreasing values of $\left|F_\ell\right|$ at higher values of $\ell$, as one can infer from Fig.~\ref{fig:cut}, top middle panel.

In contrast to the odd shells, the even shell number $\ell=342$ (cf. right panel in Fig.~\ref{fig:fourier_coefficients_N}) shows a pronounced $\Delta N = 2$ staggering.
It can be understood from the symmetry properties of $A$, $E$, $\omega$ and $\Theta$ w.r.t.\ $t=T/4$, from which $F_\ell = 0$ for $p_\parallel = 0$, $k$ even and $N$ odd follows.
In particular one field, i.e.\ $E_2=0$, causes only peaks in $f$ related to odd shell numbers.
This is already evident in the bottom middle panel in Fig.~\ref{fig:cut}, where no even shells appear at $p_\parallel=0$.
(For $p_\parallel\ne0$ however, even shells appear which may display further zeroes on $p_\perp^{(\ell)}(p_\parallel)$, see Fig.~\ref{fig:contour} in the Appendix.)
The pattern described continues to higher shell numbers, with decreasing values as for odd shells.
The widespread changes of the Fourier coefficients under variations of $N$ and $E_2$ at frozen-in values of $T$ and $E_1$ let us argue that a simple analytical formula can hardly provide an adequate description in the considered parameter range.

Having discussed the amplification effect for a variation of the fast weak field parameters $E_2$ and $\nu_2$ by means of the Fourier coefficients, let us consider variations of $E_1$ and $\nu_1$.
Keeping 50 (5+5) oscillations of field ``1'' within the flat-top (ramping+deramping) time and $(E_2/E_c,\, \nu_2/m) = (\num{0.05},\, \num{0.5})$ we make variations of $\nu_1/m$ down to \num{0.0025} at fixed $E_1/E_c=\num{0.1}$ (i.e.\ $\gamma_1=\num{0.025}$).
The spectra (calculated by means of (\ref{eq:integro}, \ref{eq:ke})) for field ``1'' alone and for fields ``1+2'' look similar to the respective panels in Fig.~\ref{fig:cut} with (i) more closely spaced peaks due to smaller $\nu_1$ and (ii) peak maxima somewhat reduced.
That means our amplification is robust, as also under variations of $E_1$ (keeping $E_1>E_2$), as confirmed by an analysis of the Fourier coefficients.

As anticipated in section~\ref{sec:shell_structure_shape}, enlarging $\tau_\text{f.t.}$ makes the peaks (shells) higher and sharper (cf.~(\ref{eq:f_on_shell}, \ref{eq:shell_profile})), while the pedestrials (accessible by~(\ref{eq:integro}, \ref{eq:ke})) hardly change.
The ramping interval $\tau_\text{ramp}$ must not be too short to avoid unwanted spikes bracketing the electric field;
larger values of $\tau_\text{ramp}$ can be accomodated in an enlarged effective $\tau_\text{f.t.}$.

Finally, we mention that non-integer values of $N$ result in a similar (albeit non-resonant) amplification, however, with a more involved phase space distribution which is no longer accessible by the harmonic analysis in section~\ref{sec:shell_structure_shape}.

\section{Summary\label{sec:summary}}
In the present work we have considered the dynamically assisted Schwinger effect for resonant periodic fields within the framework of the quantum kinetic equation.
We have isolated a non-linear parametric mechanism which increases the pair creation rate by many orders of magnitude when combining suitably a strong low-frequency field with a weak high-frequency field compared to the rates if both fields acted alone.
Both fields are subcritical with respect to frequencies and field strengths.
In contrast to previous work, which often deals with instantaneous switching off, the residual phase space distribution exhibits a distinct shell structure which survives the involved transiently oscillating pattern during the time-limited action of the periodic fields.
The occupancy of the shells grows linearly with the flat-top time, while the shell peaks grow quadratically due to a new resonance like behaviour.
The obvious motivation for such a configuration of combined two periodic fields is the superposition of the European XFEL with an ultra-intense optical laser system as envisaged in HIBEF \cite{HIBEF}.
For an easy numerical treatment, however, we have selected, in the present case study, patches in the field-strength vs.\ frequency space which, while located in the tunneling and multi-photon domains respectively, are quite different from more realistic values, for example those in table~1 in~\cite{ringwald_pair_2001}.
Based on the systematics presented here, we argue that no qualitative changes arise when moving towards parameters being more representative for an optical laser-XFEL combination.

\section*{Acknowledgments}
T.~E.~Cowan and R.~Sauerbrey are gratefully acknowledged for a fruitful collaboration within the HIBEF project at European XFEL.
The authors thank R.~Alkofer, H.~Gies, S.~S.~Schmidt, and R.~Sch\"utzhold for inspiring discussions.
The work of D.~Blaschke was supported in part by the Polish Ministry of Science and Higher Education (MNiSW) under grant no.\ 1009/S/IFT/14.

\bibliographystyle{custom}
\bibliography{lit}

\begin{thebibliography}{10}

\bibitem{schwinger}
J.~Schwinger, Phys.\ Rev. 82 (1951) 664.

\bibitem{sauter}
F.~Sauter, Z.\ Phys. 69 (1931) 742.

\bibitem{basar_magneto-sono-luminescence_2014}
G.~Basar et~al., {arXiv}:1402.2286  (2014).

\bibitem{tuchin_particle_2013}
K.~Tuchin, Adv.\ High Energy Phys. 2013 (2013) 490495.

\bibitem{kouveliotou_x-ray_1998}
C.~Kouveliotou et~al., Nature 393 (1998) 235.

\bibitem{mereghetti_strongest_2008}
S.~Mereghetti, Astron.\ Astrophys.\ Rev. 15 (2008) 225.

\bibitem{casher_chromoelectric-flux-tube_1979}
A.~Casher et~al., Phys.\ Rev.\ D 20 (1979) 179.

\bibitem{roberts_dyson-schwinger_2000}
C.~Roberts and S.~Schmidt, Prog.\ Part.\ Nucl.\ Phys 45 (2000) S1.

\bibitem{prozorkevich_vacuum_2004}
A.~V. Prozorkevich et~al., Phys.\ Lett.\ B 583 (2004) 103.

\bibitem{fedotov_limitations_2010}
A.~M. Fedotov et~al., Phys.\ Rev.\ Lett. 105 (2010) 080402.

\bibitem{bell_possibility_2008}
A.~Bell and J.~Kirk, Phys.\ Rev.\ Lett. 101 (2008) 200403.

\bibitem{elkina_qed_2011}
N.~V. Elkina et~al., Phys.\ Rev.\ {ST} Accel.\ Beams 14 (2011) 054401.

\bibitem{nerush_laser_2011}
E.~N. Nerush et~al., Phys.\ Rev.\ Lett. 106 (2011) 035001.

\bibitem{birrell_davies}
N.~D. Birrell and P.~C.~W. Davies, \emph{Quantum fields in curved spacetime},
  Cambridge University Press (1982).

\bibitem{mamayev_mostepanenko}
V.~M. Mostepanenko et~al., \emph{Vacuum Quantum Effects in Strong Fields},
  Friedmann Laboratory Publishing Ltd., St.~Petersburg (1994).

\bibitem{brezin_pair_1970}
E.~Brezin and C.~Itzykson, Phys.\ Rev.\ D 2 (1970) 1191.

\bibitem{blaschke_properties_2013}
D.~B. Blaschke et~al., Phys.\ Rev.\ D 88 (2013) 045017.

\bibitem{narozhny_pair_2004}
N.~B. Narozhny et~al., Phys.\ Lett.\ A 330 (2004) 1.

\bibitem{ELI}
{Proposal} for a {European} {Extreme} {Light} {Infrastructure} ({ELI}),
  \url{http://www.extreme-light-infrastructure.eu/pictures/ELI-scientific-case-id17.pdf}.

\bibitem{dreisow_vacuum_2012}
F.~Dreisow et~al., Phys.\ Rev.\ Lett. 109 (2012) 110401.

\bibitem{schutzhold_dynamically_2008}
R.~Sch{\"u}tzhold et~al., Phys.\ Rev.\ Lett. 101 (2008) 130404.

\bibitem{fey_momentum_2012}
C.~Fey and R.~Sch{\"u}tzhold, Phys.\ Rev.\ D 85 (2012) 025004.

\bibitem{dunne_catalysis_2009}
G.~V. Dunne et~al., Phys.\ Rev.\ D 80 (2009) 111301.

\bibitem{orthaber_momentum_2011}
M.~Orthaber et~al., Phys.\ Lett.\ B 698 (2011) 80.

\bibitem{kohlfurst_optimizing_2013}
C.~Kohlf{\"u}rst et~al., Phys.\ Rev.\ D 88 (2013) 045028.

\bibitem{sicking_bachelor_2012}
J.~Sicking, \emph{{Puls\-form\-abh\"angig\-keit} im dynamisch
  ver\-st{\"a}rk\-ten {Sauter}-{Schwinger}-{Effekt}}, Bachelor thesis,
  Universit{\"a}t Duisburg-Essen (2012).

\bibitem{hebenstreit_optimization_2014}
F.~Hebenstreit and F.~Fillion-Gourdeau, Phys.\ Lett.\ B 739 (2014) 189.

\bibitem{akal_electron-positron_2014}
I.~Akal et~al., {arXiv}:1409.1806  (2014).

\bibitem{ringwald_pair_2001}
A.~Ringwald, Phys.\ Lett.\ B 510 (2001) 107.

\bibitem{nuriman_enhanced_2012}
A.~Nuriman et~al., Phys.\ Lett.\ B 717 (2012) 465.

\bibitem{oluk_electron-positron_2014}
O.~Oluk et~al., Front.\ Phys. 9 (2014) 157.

\bibitem{schmidt_quantum_1998}
S.~M. Schmidt et~al., Int.\ J.\ Mod.\ Phys.\ E 7 (1998) 709.

\bibitem{blaschke_pair_2006}
D.~Blaschke et~al., Phys.\ Rev.\ Lett. 96 (2006).

\bibitem{blaschke_influence_2013}
D.~B. Blaschke et~al., Contrib.\ Plasm.\ Phys. 53 (2013) 165.

\bibitem{smolyansky_photon_2012}
S.~A. Smolyansky et~al., PoS (Baldin ISHEPP XXI)  (2012) 069.

\bibitem{kleinert_electron-positron_2008}
H.~Kleinert et~al., Phys.\ Rev.\ D 78 (2008) 025011.

\bibitem{di_piazza_pair_2004}
A.~D. Piazza, Phys.\ Rev.\ D 70 (2004) 053013.

\bibitem{gies_pair_2005}
H.~Gies and K.~Klingm{\"u}ller, Phys.\ Rev.\ D 72 (2005) 065001.

\bibitem{dabrowski_super-adiabatic_2014}
R.~Dabrowski and G.~V. Dunne, {arXiv}:1405.0302  (2014).

\bibitem{hebenstreit_diss_2011}
F.~Hebenstreit, \emph{Schwinger effect in inhomogeneous electric fields}, Ph.D.
  thesis, Karl-Franzens-Universit{\"a}t Graz (2011).

\bibitem{alkofer_pair_2001}
R.~Alkofer et~al., Phys.\ Rev.\ Lett. 87 (2001) 193902.

\bibitem{mocken_nonperturbative_2010}
G.~R. Mocken et~al., Phys.\ Rev.\ A 81 (2010) 022122.

\bibitem{ruf_pair_2009}
M.~Ruf et~al., Phys.\ Rev.\ Lett. 102 (2009) 080402.

\bibitem{popov_resonant_1973}
V.~S. Popov, {JETP} Lett. 18 (1973) 255.

\bibitem{narozhny_pair_1973}
V.~S. Popov and A.~I. Nikishov, Sov.\ Phys.\ {JETP} 38 (1974) 427.

\bibitem{mostepanenko_1974}
V.~M. Mostepanenko and V.~M. Frolov, Sov.\ J.\ Nucl.\ Phys. 19 (1974) 451.

\bibitem{popov_1974}
V.~S. Popov, Sov.\ J.\ Nucl.\ Phys. 19 (1974) 584.

\bibitem{hebenstreit_momentum_2009}
F.~Hebenstreit et~al., Phys.\ Rev.\ Lett. 102 (2009) 150404.

\bibitem{jiang_enhanced_2013}
M.~Jiang et~al., Chin.\ Phys.\ B 22 (2013) 100307.

\bibitem{ren_pair_2012}
N.~Ren et~al., Chin.\ Phys.\ Lett. 29 (2012) 071201.

\bibitem{schmidt_non-markovian_1999}
S.~Schmidt et~al., Physical Review D 59 (1999) 094005.

\bibitem{kohlfurst_effective_2014}
C.~Kohlf{\"u}rst et~al., Phys.\ Rev.\ Lett. 112 (2014) 050402.

\bibitem{HIBEF}
The {HIBEF} project, \url{www.hzdr.de/hgfbeamline/}.

\end{thebibliography}

\newpage
\appendix
\setcounter{figure}{0}
\renewcommand{\thefigure}{A\arabic{figure}}
\section*{Appendix: Phase space distributions}
\begin{sidewaysfigure}
\centering
\includegraphics[width=\textwidth]{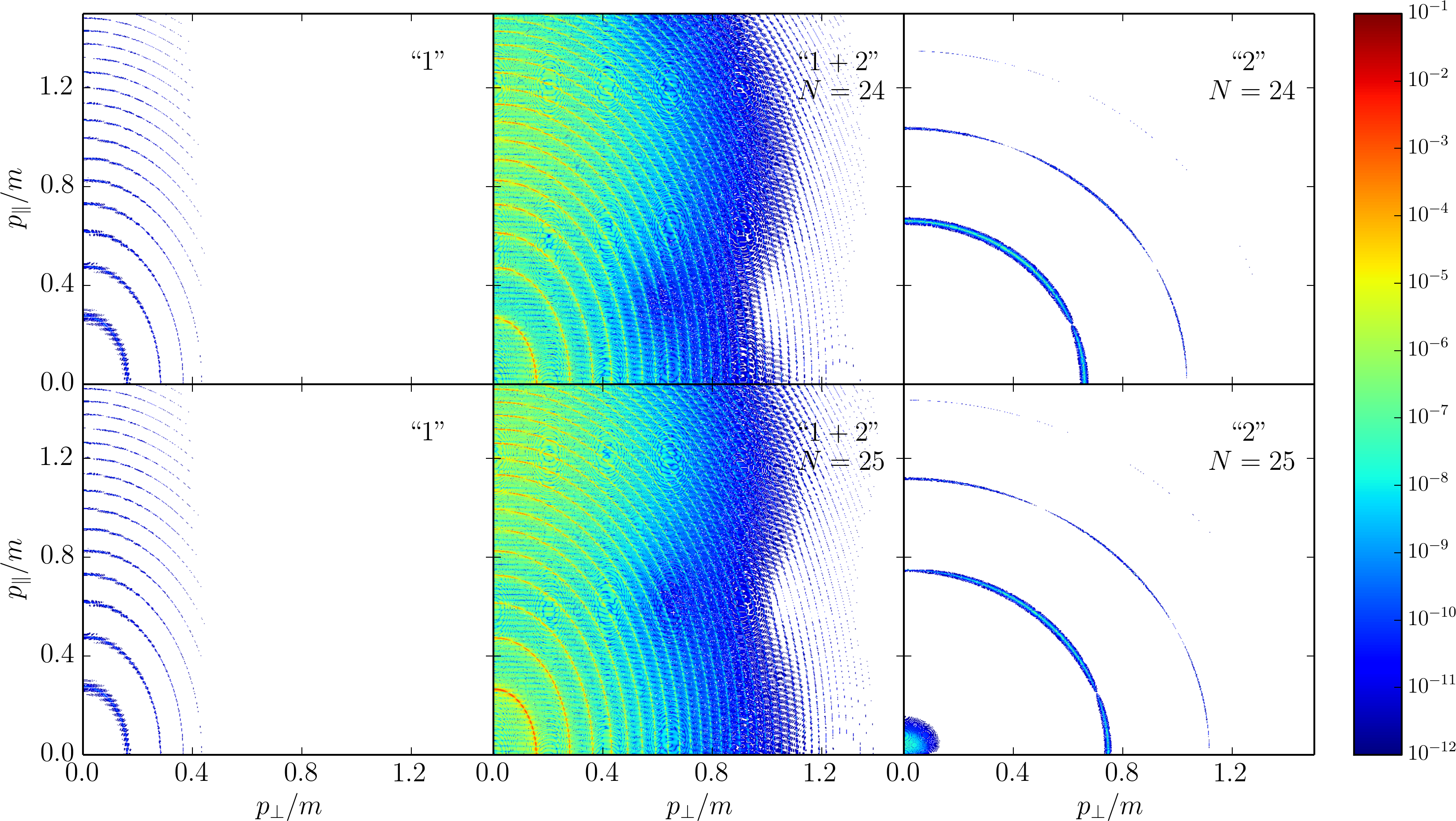}
\caption{(color online) As Fig.~\ref{fig:cut} but for $f$ as a function of $p_\perp$ and $p_\parallel$ with the color code for $f$ on the right.}
\label{fig:contour}
\end{sidewaysfigure}

We show in Fig.~\ref{fig:contour} plots of $f$ over the full phase space, i.e.~the distribution over the $p_\perp$-$p_\parallel$ plane.
Figure~\ref{fig:cut} is a cross section of these contour plots at $p_\parallel=0$.
These plots unravel fairly rich structures along the ridges, such as deep notches (the missing $\ell=\text{even}$ peaks at $p_\parallel=0$ in Fig.~\ref{fig:cut} are a consequence), the steeper dropping of the ridge maximum in $p_\perp$ direction, and the degree of anisotropy (that is the elongation in $p_\parallel$ direction).
The ridge structure is nevertheless well described by~\eqref{eq:shell_condition};
some details are uncovered by generalizing~(\ref{eq:f_on_shell}, \ref{eq:shell_profile}) to non-zero $p_\parallel$.
These pecularities of the full phase space distribution are not included in the estimator formula~\eqref{eq:density_estimator}.
Instead, it is meant to expose the rough dependence on the Fourier coefficients~\eqref{eq:fourier_coefficients} and to deliver an order of magnitude orientation.
In fact, comparing the densities $n$ in units of $m^3$ from~\eqref{eq:density_estimator} with a numerical evaluation (num.\ eva.) we find\\[1ex]
\begin{tabular}{ll|ccc}
                        &                              & ``1''       & ``1+2''    & ``2''\\\hline
\multirow{2}{*}{$N=24$} & num.\ eva.                   & \num{2e-12} & \num{7e-6} & \num{2e-8}\\
                        & \eqref{eq:density_estimator} & \num{3e-13} & \num{1e-6} & \num{1e-8}\\\hline
\multirow{2}{*}{$N=25$} & num.\ eva.                   & \num{2e-12} & \num{1e-5} & \num{6e-9}\\
                        & \eqref{eq:density_estimator} & \num{3e-13} & \num{4e-6} & \num{8e-9}
\end{tabular}\\[1ex]
showing that~\eqref{eq:density_estimator} must be employed with care.
\end{document}